\documentclass[aps,pra,showpacs,twocolumn,superscriptaddress]{revtex4-1}
\usepackage{amsmath}
\usepackage{amsfonts,bbold}
\usepackage{amssymb}
\usepackage{bbm,bm}
\usepackage{graphicx}
\usepackage{natbib}
\usepackage[colorlinks=true,linkcolor=blue,urlcolor=black,citecolor=blue]{hyperref}

\begin{document}

\title{Optically-mediated remote entanglement generation in magnon-cavity systems}

\author{Da-Wei Luo}
\affiliation{Center for Quantum Science and Engineering and Department of Physics, Stevens Institute of Technology, Hoboken, New Jersey 07030, USA}

\author{Xiao-Feng Qian}
\affiliation{Center for Quantum Science and Engineering and Department of Physics, Stevens Institute of Technology, Hoboken, New Jersey 07030, USA}

\author{Ting Yu}
\email{Corresponding Author: Ting.Yu@stevens.edu}
\affiliation{Center for Quantum Science and Engineering and Department of Physics, Stevens Institute of Technology, Hoboken, New Jersey 07030, USA}

\date{\today}

\begin{abstract}

We study the remote entanglement generation between macroscopic microwave magnon modes in a coupled cavity system. The cavities are connected via an optical fiber, which necessitates the use of a frequency conversion inside the cavity. The converter may be implemented via a rare-earth doped crystal acting like an effective three-level system. The entanglement dynamics of the system is analytically studied, and an active optimal control method is also proposed where one may generate maximally entangled Bell states on demand with a given evolution time. The system dynamics and its control have also been studied in a generic non-Markovian open system framework, and the generated entanglement is found to be robust against environmental noises.

\end{abstract}

\maketitle

\section{Introduction}

Quantum magnon systems are a rapidly developing research field that has been studied in many contexts related to quantum information applications in recent years. Experiments
with the magnon systems have been explored in the cavity-magnon interaction, entanglement generation, and control~\cite{Rameshti2022j,Azimi-Mousolou2021p,Luo2021q,Rao2021y,Etesamirad2021d}. It has been found that macroscopic entanglement can be established in
the magnon modes, which is typically found in magnetic materials such as yttrium iron garnet (YIG) spheres~\cite{Tabuchi2014a,you_nat1,Lachance-Quirion2017a}, and has also been suggested to reside in Bose-Einstein condensates~\cite{Nikuni2000q}. The magnon system has shown itself to be a versatile quantum entity that can be coherently coupled to superconducting qubits~\cite{Tabuchi2015a,Luo2021q,Tabuchi2016a}, optical cavities~\cite{Wang2018a,you_nat1,you_nat2}, and the mechanical vibration mode of the YIG sphere~\cite{Jie-Li2021x}. Such versatility makes the system a useful quantum information carrier that can be incorporated into various existing information processing platforms in quantum optics and other experimentally available setups. Various measurement strategies for magnons have also been reported, such as the magnon polarization~\cite{Nambu2020h} and nonlinear foldover effect~\cite{Hyde2018v} which can be directly measured.

 The magnon systems have also been used in quantum metrology such as probes for measurements of magnetic fields~\cite{Crescini2021r}, dark matter~\cite{Trickle2020u} and the gravitational wave~\cite{Ito2020q}. It is therefore of great interest to study the generation of entanglement between magnons~\cite{Yuan2020n,Luo2021q}, where it may then be used for quantum metrology tasks to achieve high precision measurements and weak signal detections. To potentially form an entangled detector array, it is necessary to study the remote coupling and entanglement generation among a set of magnon systems. One promising strategy is to link the cavity magnon system via optical fibers. However, it is known that the magnon modes are typically in the microwave range and are orders-of-magnitude smaller than the optic frequencies supported by current optical fiber components. Hence, some frequency conversion is necessary in this setup. Various frequency conversion protocols have been proposed, such as using non-linear quantum optics effects~\cite{Brecht2011s,Huang2013o} or rare-earth doped materials with external pumps~\cite{conv_ion3,conv_ion1,conv_ion2}. Such frequency converters may be used to couple cavity magnon systems together with conventional optical fibers.

In this paper, we will study the entanglement generation in a remote cavity-magnon system, where the cavity has both a microwave mode and an optical mode. The microwave mode of the cavity is coupled to a magnon mode and the frequency converter inside the cavity would introduce an effective coupling between the cavity's microwave and optical mode. The optical modes of the two remote cavities can then be coupled to each other through an optical fiber. The entanglement dynamics as well as its active control would be studied, and the influence of open system effects will also be fully considered in a generic non-Markovian framework. 

This paper is organized as follows. We first introduce the coupled magnon cavity system under consideration, and solve for its dynamics and study the behaviors of entanglement generation in the system in section~\ref{sec_sys}. An optimal control strategy would be introduced in section~\ref{sec_ctrl} where the generation of the maximally entangled Bell state is studied. In practical scenarios, the quantum systems are inevitably coupled to their environments and may be susceptible to the noises induced by the open system effects. The open system effects on the dynamics as well as the optimal control would be investigated in section~\ref{sec_opens} with a generic non-Markovian treatment. We conclude with some discussions and remarks in section~\ref{sec_conclu}, and some details of the calculation are shown in the Appendix.

%%%%%%%%%%%%%%
\begin{figure}
 \centering
 \includegraphics[width=.45\textwidth]{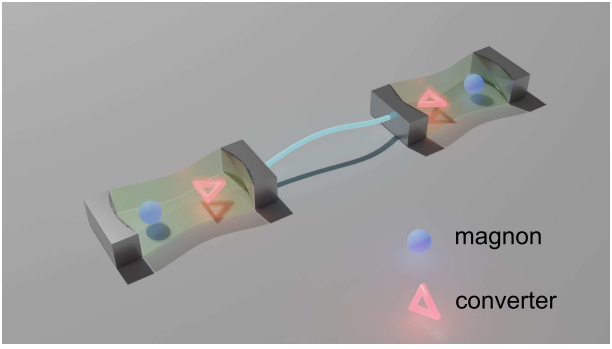}
 \caption{Schematics of the model under consideration. The cavities are dual-mode supporting both optical and microwave frequencies. Inside the cavities there is a magnetic sphere (e.g. YIG) that houses a magnon mode and a frequency converter composed of rare-earth doped crystals. The frequency converter can transfer the excitation in the microwave frequency to the optical frequency. The optical modes of the cavities can be coupled with an optical fiber.}\label{fig_model}
\end{figure}

\section{Model and solution}\label{sec_sys}

\begin{figure}
 \centering
 \includegraphics[width=.45\textwidth]{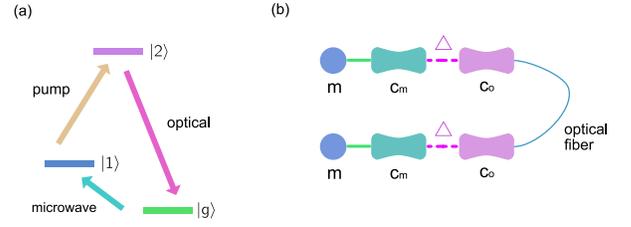}
 \caption{(a) Schematic for the energy levels of the frequency converter, which takes in a microwave frequency and converts it into the optical frequency with an external pump. (b) Effective coupling of the model under consideration: the magnon $m$ couples directly to the microwave mode $c_m$ of the cavity, whereas the effective coupling between the microwave and optical mode $c_o$ of the cavity is mediated by the frequency converter $\triangle$. The optical modes are then coupled with an optical fiber.}\label{fig_lvl}
\end{figure}

For the remote entanglement generation, we consider two cavity-magnon systems that are coupled to each other through an optical fiber. It is known that the magnon modes operate under microwave frequencies, therefore, one needs a physical setting that can convert 
the microwave frequencies to optical frequencies. One possible realization is made by carrying out the frequency conversion outside of the cavity magnon system at both ends which introduces an effective coupling between the microwave cavities~\cite{Luo2021q}. Alternatively, one may convert the frequencies within the cavity-magnon system using a dual-mode cavity that supports a microwave mode (in tune with the magnon mode) as well as an optical mode. Inside the cavity, there would be a magnetic sphere housing a magnon mode, and a frequency conversion apparatus, which may be implemented as a rare-earth doped crystal~\cite{conv_ion3,conv_ion1,conv_ion2}. The optical modes of the cavities may then be coupled with an optical fiber. A schematic representation of the model is displayed in Fig.~\ref{fig_model}. The rare-earth doped crystal~\cite{conv_ion3,conv_ion1,conv_ion2} effectively acts like a three-level system, where the microwave excitation would drive it from the ground state $|g \rangle$ to $|1 \rangle$, and an external laser subsequently pump it to the $|2 \rangle$ state. The emission of the transition from $|2 \rangle$ to the ground state $|g \rangle$ yields the required optical frequency. The scheme for this conversion is shown in Fig.~\ref{fig_lvl} (a).

The full Hamiltonian for the system is given by

\begin{align}
 H &= \sum_{i=1,2}\left[ \omega_a' a_i ^\dagger a_i + \omega_b' b_i ^\dagger b_i + \omega_m m_i ^\dagger m_i \nonumber \right. \\
 & + \delta_2 \sigma_{22,i} + \delta_1 \sigma_{11,i} + g_{mb} (m_i ^\dagger b_i + m_i b_i ^\dagger) + \Omega(\sigma_{12,i}+\sigma_{21,i}) \nonumber \\
 & \left. + g_{cb}(\sigma_{1g,i}b_i + \sigma_{g1,i} b_i ^\dagger) + g_{ca}(\sigma_{2g,i}a_i + \sigma_{g2,i} a_i ^\dagger) \right] \nonumber \\
 & + j_a(a_1 ^\dagger a_2 + a_1 a_2 ^\dagger),\label{eq_h0}
\end{align}
where $a_i$ is the optical cavity mode of the $i$th magnon cavity system with frequency $\omega'_a$, $b_i$ is the microwave cavity mode with frequency $\omega'_b$, $m_i$ the magnon mode with frequency $\omega_m$, and $\sigma_{kj,i}=|k \rangle\langle j|$ for the three-level conversion atom in the $i$-th cavity ($i=1, 2$), $\Omega$ is the pump frequency, $\delta_{1(2)}$ is the energy of the first (second) excited level of the converter, $g_{mb}$ denotes the interaction strength between the magnon and the microwave mode, $g_{cb(ca)}$ denotes the coupling between the microwave (optical) cavity mode and the converter, and $j_a$ is the fiber coupling strength. Note that the rotating wave approximation (RWA) has been applied in Eq.~\ref{eq_h0}, which generally requires the coupling strength to be much smaller than the cavity and atom frequency. However, our numerical analysis shows that, with the state space and initial condition considered in this paper, the RWA is still a viable treatment and gives accurate results for the magnon-magnon entanglement dynamics even when the coupling strength is comparable to the order of the cavity frequency. In addition, the RWA analysis has the advantage of allowing for full analytical derivations and long-time stable simulations. Therefore, we will adopt this RWA treatment throughout this work.

The difference between the optical and microwave frequencies results in a large-detuning situation, which consequently allows one to adiabatically eliminate~\cite{Lugiato2015j,Brion2007k} the $|2 \rangle$ and $|1 \rangle$ levels of the three-level converter. As such, we have an effective Hamiltonian for the dual-mode cavity magnon system as
\begin{align}
 H_{\rm eff} &= \sum_{i=1,2}\left[ \omega_a a_i ^\dagger a_i + \omega_b b_i ^\dagger b_i + \omega_m m_i ^\dagger m_i \nonumber \right. \\
 & \left. + g_m (m_i ^\dagger b_i + m_i b_i ^\dagger) + g_{c}(a_i b_i^\dagger + a_i^\dagger b_i) \right] \nonumber \\
 & + j_a(a_1 ^\dagger a_2 + a_1 a_2 ^\dagger),\label{eq_heff}
\end{align}
where $\omega_a$ is the optical mode frequency, $\omega_b$ is the microwave mode frequency, $\omega_m$ is the frequency of the magnon mode, $g_m$ is the coupling between the magnon and microwave mode, $j_a$ is the coupling strength of the optical fiber between the two optical modes of the cavity and the effective coupling between the optical and microwave mode of the cavity is given by
$g_c = \frac{g_{ca} g_{cb} \Omega}{\delta_2 \delta_3-\Omega ^2}$.
An illustration for this effective Hamiltonian is shown in Fig.~\ref{fig_lvl} (b), which has a chain-like structure.

\begin{figure}[t]
 \centering
 \includegraphics[width=.47\textwidth]{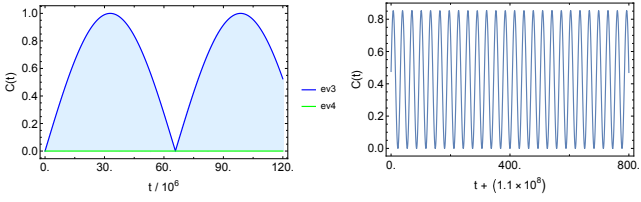}
 \caption{Entanglement of the two magnon modes as a function of time. (Left panel) The entanglement dynamics is fast oscillating and exhibits a beat pattern due to the superposition of oscillations of slightly different frequencies. An approximation of the beat's envelope can also be obtained analytically, given by $ev3$ and $ev4$ in the appendix. (Right panel) A zoom-in of $t \in [110000000,110000800]$, showing the fast oscillations of the entanglement dynamics.}\label{fig_cm_e}
\end{figure}

Note that the effective Hamiltonian Eq.~\eqref{eq_heff} conserves the total number of excitations in the system. Let us consider the dynamics of the system in the single-excitation subspace, where an initial excitation is in the magnon mode of the first magnon-cavity system. This treatment allows us to use the magnon as an effective two-level system, which is useful in quantum information processing tasks. The initial state of the system can therefore be written as $|\psi(0) \rangle = |100000 \rangle$ where the basis is $|m_1,m_2,c_{m1},c_{m2},c_{o1},c_{o2} \rangle$ for excitation in the first (second) magnon, microwave mode and optical mode, respectively. The entanglement dynamics measured by the concurrence~\cite{Wootters1998a} is shown in Fig.~\ref{fig_cm_e}, taking $\omega_b=\omega_m=1$, $\omega_a=1200$, $g_m=0.1$, $g_c=0.23$ and $j_a=1.3$. It can be observed that the concurrence dynamics is fast-oscillating and shows a periodic envelope. The analytical expressions for the envelope may also be analytically derived (see Appendix~\ref{sec_apnx}).

Numerical and analytical solutions to our system may be obtained by using a Feshbach P-Q partition technique~\cite{Wu2009,Jing2014,Luo2016s}. In the basis where there is one excitation in the combined system including magnons, microwave cavities, optical cavities $m_i,c_{m,i},c_{o,i}$, \(i=1,2\), the Hamiltonian takes a more concrete matrix form,
\begin{align}
 H' = 
 \begin{bmatrix}
 \omega_m & 0 & g_m & 0 & 0 & 0 \\
 0 & \omega_m & 0 & g_m & 0 & 0 \\
 g_m & 0 & \omega_b & 0 & g_c & 0 \\
 0 & g_m & 0 & \omega_b & 0 & g_c \\
 0 & 0 & g_c & 0 & \omega_a & j_a \\
 0 & 0 & 0 & g_c & j_a & \omega_a \\
 \end{bmatrix}
\end{align}

Since the optical frequency $\omega_a$ is orders-of-magnitude larger other parameters, we consider $\omega_m=\omega_b$ and first rotate out the optical part
\begin{align}
 \mathrm{diag}\left(0_{4},\begin{bmatrix}
 \omega_a & j_a \\
 j_a & \omega_a \\
 \end{bmatrix}\right),
\end{align}
where $0_4$ is a $4\times 4$ matrix whose elements are $0$, we have
\begin{align}
 H_I = \begin{bmatrix}
 \omega_m & 0 & g_m & 0 & 0 & 0 \\
 0 & \omega_m & 0 & g_m & 0 & 0 \\
 g_m & 0 & \omega_m & 0 & g_{cc}^* & -i g_{cs}^* \\
 0 & g_m & 0 & \omega_m & -i g_{cs}^* & g_{cc}^* \\
 0 & 0 & g_{cc} & i g_{cs} & 0 & 0 \\
 0 & 0 & i g_{cs} & g_{cc} & 0 & 0 \\
 \end{bmatrix},
\end{align}
where $g_{cc}=g_c e^{i \omega_a t} \cos (j_a t)$, $g_{cs}=g_c e^{i \omega_a t} \sin (j_a t)$ are the fast-oscillating elements.

Take \(\mathcal{P}=\sum_{i=1,2} |m_i \rangle\langle m_i| + |c_{mi} \rangle\langle c_{mi}| =\mathrm{diag}(1_4,0_2) \) and \(\mathcal{Q}=\sum_{i=1,2}|c_{oi} \rangle\langle c_{oi}|=\mathbbm{1}-\mathcal{P}=\mathrm{diag}(0_4,1_2)\), where $1_n$ is an identity matrix of size $n\times n$, and let $\mathcal{P}|\varphi \rangle = p$, $\mathcal{Q}|\varphi \rangle = q$ we have

\begin{align*}
 \partial_t \mathcal{P}|\varphi \rangle &= -i \mathcal{P} H_I(t) \mathcal{P}|\varphi \rangle-i \mathcal{P} H_I(t) \mathcal{Q}|\varphi \rangle \\
 &= -i h \mathcal{P}|\varphi \rangle-i R \mathcal{Q}|\varphi \rangle
\end{align*}
which can be denoted as
\begin{align}
 i\partial_t p = h p + R q. \label{eq_dp}
\end{align}
Similarly,
\begin{align*}
\partial_t \mathcal{Q}|\varphi \rangle&=-i \mathcal{Q} H_I(t) \mathcal{P}|\varphi \rangle-i \mathcal{Q} H_I(t) \mathcal{Q}|\varphi \rangle \\
 &=-i W \mathcal{P}|\varphi \rangle-i D \mathcal{Q}|\varphi \rangle
\end{align*}
and again can be denoted as
\begin{align}
 i \partial_t q &= W p + D q.
\end{align}
Since $q(0)=0$ and $D=0$, we have, formally, $q=-i\int ds W(s)p(s)$. Inserting this into Eq.~\eqref{eq_dp}, we finally get an integral-differential equation
\begin{align}
 i \partial_t p &= h p - i R \int ds W(s)p(s) \label{eq_pq_dp}
\end{align}

where
\begin{align}
 W(s)p(s) &= \left[
 \begin{array}{c}
 g_c e^{i \omega_a s} [p_3(s) \cos (j_a s) + i p_4(s) \sin (j_a s)] \\
 g_c e^{i \omega_a s} [p_4(s) \cos (j_a s) + i p_3(s) \sin (j_a s)] \\
 \end{array}
 \right],\label{eq_wsps}
\end{align}
where $p_i$ denotes the $i$-th component of the wave function in the $\mathcal{P}$ projected space. Eq.~\eqref{eq_pq_dp} is difficult to solve directly since it's an integral-differential equation. Taking into consideration that $\omega_a \gg \omega_m, g_m, g_c, j_a$, we may convert Eq.~\eqref{eq_pq_dp} into an ordinary differential equation. This is possible since the integrand Eq.~\eqref{eq_wsps} is of the form $f(s) \exp(i \omega_a s)$ where $f(s)$ is a generic function, we may approximate the term $\int ds W(s)p(s)$ using integral-by-parts to express it in powers of $1/\omega_a$: let $I$ be
\begin{align}
 I&= \int_a^b f(\tau)\exp\left(i \omega_a \tau\right) d\tau \nonumber \\
 &=\epsilon \left[-i f(\tau)\exp\left(i \omega_a \tau\right)\right]|_{\tau=a}^b \nonumber \\
 & -\epsilon \int_a^b \frac{d}{d\tau}\left[-i f(\tau)\right] \exp\left(i \omega_a \tau\right) d\tau, \label{eq_intsim}
\end{align}
where $\epsilon=1/\omega_a$ and the second term is just $I$ with $f \rightarrow \frac{d}{d\tau}\left[-i f(\tau)\right]$ and contributes to the higher order $O(\epsilon^2)$. Repeatedly using this formula would enable one to write $I$ in powers of $\epsilon$. We may now apply Eq.~\eqref{eq_intsim} to~\eqref{eq_wsps}
and keep up to $O(\epsilon^3)$ (since $O(\epsilon^2)$ cancels out fiber coupling), we have

\begin{align}
 i \partial_t p = h_{\rm eff} p + p_d,
\end{align}
where $h_{\rm eff}$ is an effective non-Hermitian Hamiltonian with $O(\epsilon^2)$ corrections (due to the optical frequency), given by 
\begin{widetext}
\begin{align}
 h_{\rm eff}
 &= \left(
 \begin{array}{cccc}
 \omega_m & 0 & g_m & 0 \\
 0 & \omega_m & 0 & g_m \\
 g_m-\frac{g_c^2 g_m}{\omega_a^2} & 0 & \frac{-g_c^2 (\omega_m+\omega_a)}{\omega_a^2}+\omega_m & \frac{g_c^2 j_a}{\omega_a^2} \\
 0 & g_m-\frac{g_c^2 g_m}{\omega_a^2} & \frac{g_c^2 j_a}{\omega_a^2} & \frac{-g_c^2 (\omega_m+\omega_a)}{\omega_a^2}+\omega_m \\
 \end{array}
 \right).
\end{align}
\end{widetext}
and the drift term $p_d$ is on the order of $O(\epsilon^2)$ is discarded. 

The wave function components for the two magnons, up to $O(\epsilon)$, may then be given by
\begin{align}
 \psi_{m1} &= p_1(t) = \left(e^{i\omega_1t} + e^{-i\omega_2t} + e^{i\omega_3t} + e^{-i\omega_4t} \right)/4 \nonumber \\
 \psi_{m2} &= p_2(t) = \left(e^{i\omega_1t} + e^{-i\omega_2t} - e^{i\omega_3t} - e^{-i\omega_4t} \right)/4,
\end{align}
where
\begin{align}
 \omega_{1(2)} &= \left[ \pm g_c^2 (\omega_m+\omega_a-j_a) +s_1 \mp 2 \omega_m \omega_a^2 \right]/2\omega_a^2 \nonumber \\
 \omega_{3(4)} &= \left[ \pm g_c^2 (\omega_m+\omega_a+j_a) +s_2 \mp 2 \omega_m \omega_a^2 \right]/2\omega_a^2
\end{align}
and
\begin{align}
 s_{1(2)} &= \sqrt{g_c^4 (\mp j_a+\omega_m+\omega_a)^2-4 g_c^2 g_m^2 \omega_a^2+4 g_m^2 \omega_a^4}.\nonumber
\end{align}
We may now see that the wave function components corresponding to the magnons can be approximated by a superposition of oscillations with slightly different frequencies on the order of $O(1/\omega_a^2)$. The difference in frequency is induced by the large optical frequency which is orders-of-magnitude larger than other parameters, giving rise to a beat pattern with a periodic envelope. The analytical expression for the envelope may also be derived (see Appendix). In the situation where a high-resolution measurement of the entanglement or system dynamics is difficult but the envelope of the system dynamics is more feasible to measure, one may instead deduce the system parameters from the envelopes. This feature may be useful in situations where metrology of the system parameters is used in the detection of weak signals, where the signal alters the system parameters.

\section{Controlled entanglement generation}\label{sec_ctrl}

The entanglement dynamics discussed in the previous section has shown a rapid oscillatory behavior due to the difference between the microwave and optical frequencies. In some situations, the remote entanglement may be needed at the prescribed time $T$. In such a situation, it is desirable to explore how to generate a maximally entangled state at a given time through an external control mechanism.

\begin{figure}
    \centering
    \includegraphics[width=.48\textwidth]{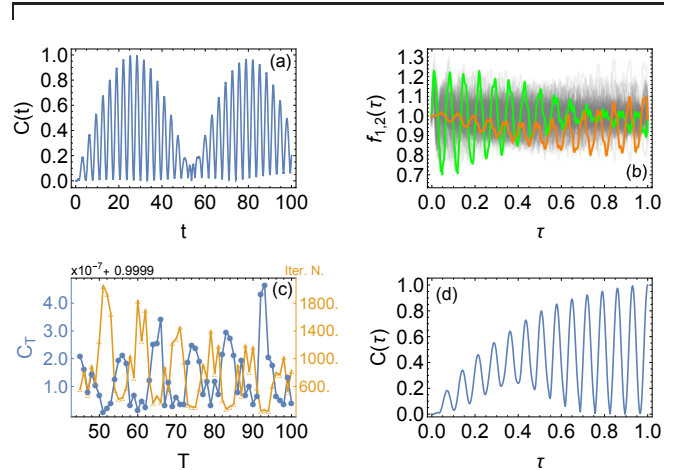}
    \caption{(Panel a). The entanglement dynamics of magnon modes without an active control. (Panel b). The control functions $f_{1(2)}(\tau)$ as functions of the scaled time $\tau=t/T$, where $T$ is the total runtime. The green (orange) line shows an example of $f_{1(2)}$ for $T=45$, and the grey lines corresponds to the control functions $f_{1(2)}$ under $T=46, 47, \ldots 100$. It can be seen that they can be kept in a range of $1\pm 0.3$. (Panel c). The entanglement of the magnons at $t=T$ for integers $T\in[45, 100]$ (blue line marked with dots). It can be seen that they are close to the maximally entangled Bell state, with concurrence $C_T > 0.9999$. The orange line marked with $\triangle$ shows the number of iterations needed to reach the target state. (Panel d.) An example of the entanglement dynamics for $T=45$, as a function of the scaled time $\tau=t/T$.}\label{fig_ctrl}
\end{figure}

For this purpose, we may choose to modulate the frequency of the magnon modes separately. Let the now time-dependent frequency $\omega_m$ for the first (second) magnon be $f_{1(2)}(t)$. The system Hamiltonian is now given by Eq.~\eqref{eq_heff} where the $\sum_{i=1,2} \omega_m m_i ^\dagger m_i$ term is replaced by $\sum_{i=1,2} f_i(t) m_i ^\dagger m_i$. The target state is set to the maximally entangled Bell state for the magnons and $|0 \rangle$ for all cavity modes, 

\begin{align}
 |\varphi_T \rangle = \frac{1}{\sqrt{2}}[|10 \rangle+|01 \rangle]_{m_1m_2} \otimes |0000 \rangle_{c_{m1}c_{m2}c_{o1}c_{o2}}.
\end{align}

The optimal shape of this modulated frequency may be obtained by a gradient-based search algorithm known as the Krotov method~\cite{Tannor1992h,Goerz2019w,Reich2012v,Konnov1999b}, which can find an optimal set of control functions $f_i(t)$ such that the Hamiltonian $H = H_0 + \sum_i f_i(t) H_i$ can drive the system from some initial state $|\psi(0) \rangle$ into a given target state $|\varphi_T \rangle$ for an evolution time of $T$. One problem faced by the quantum feedback or closed loop control is that the control function is determined by the quantum state, while on the other hand, the quantum state is also dependent on the control function through time propagation. The Krotov method is formulated with a unique approach to how it separates the interdependence between the state evolution and the control fields. The Krotov method works as an iterative algorithm where the control functions obtained from the previous iteration are used as the guess for the next one, provided with an initial guess of the controls. The goal of the Krotov method is to minimize the functional $J$ defined below,
\begin{align}
 J\left[s, \{f^{(i)}_l(t)\}\right] = J_T(s) + \sum_l \int_0^T g(\{f^{(i)}_l(t)\}) \label{eq_j}
\end{align}
where $s=\{| \psi^{(i)}(t)\}$ denotes the wave functions under the $i$-th iteration, and $\{f^{(i)}_l(t)\}$ are the control functions. In the continuos time limit~\cite{Tannor1992h,Reich2012v,Konnov1999b}, the algorithm is guaranteed to monotonically minimize the functional~\eqref{eq_j}. 

For the control problem of state engineering, $J_T$ can be set to be the infidelity between the evolved state and the target state,
\begin{align}
 J_T(s) = 1 - |\langle \varphi_{T} | \psi^{(i)}(T) \rangle|^2
\end{align}
for the $i$-th iteration, where $|\varphi_{T} \rangle$ is the target final state. The function $g$ tracks the running cost of the control fields' changes, and is usually taken in the form of
\begin{align}
 g(\{f^{(i)}_l(t)\}) = \frac{\lambda_{a,l}}{S_l(t)}(\Delta f_l^{(i)}(t))^2,
\end{align}
where $\lambda_{a,l}>0$ is an inverse step-size, $\Delta f_l^{(i)}(t)= f_l^{(i)}(t) - f_l^{\rm ref}(t)$ is the difference of the control function between the current control field and some reference control field $f_{l}^{\rm ref}(t)$, generally taken as the control functions obtained from the last iteration $f_{l}^{\rm ref}(t) \leftarrow f_l^{(i-1)}(t)$, and $S_l(t) \in [0,1]$ is the weight or update shape function. One would then start with a trial solution to the control functions, and in the $i$-th iteration, the $l$-th control field is updated according to
\begin{align}
 \Delta f_l^{(i)}(t) = \frac{S_l(t)}{\lambda_{a,l}} \mathrm{Im} \left[\left\langle \chi^{(i-1)}(t)\left|\frac{\partial H^{(i)}}{\partial f^{(i)}_l(t)} \right| \psi^{(i)}(t)\right\rangle\right],
\end{align}
where $H^{(i)}$ is the total Hamiltonian of the $i$-th iteration and $|\chi^{(i-1)}(t) \rangle$ is backwards-propagated using the Hamiltonian under the previous iteration's control fields, with an appropriate boundary condition $|\chi^{(i-1)}(T) \rangle \propto |\varphi_T \rangle$, i.e. the target state.

Here, we take the initial guesses $f_1(t)=f_2(t)=1$ and apply the Krotov method to find the controls $f_{1(2)}$, with a goal of reaching $J=1\times10^{-4}$ or the controls functions falling out of the range of $1\pm 0.3$ so that the modulated frequency does not fall out of some typical magnon mode frequency. Taking $\omega_b=1$, $\omega_a=12$, $j_a=3$, $g_m=1$ and $g_c=1.5$, we first show the uncontrolled entanglement dynamics in Fig.~\ref{fig_ctrl} (a), where we can see the fast oscillation and the beat pattern envelopes. In Fig.~\ref{fig_ctrl} (b), the optimized control $f_{1(2)}$ for $T=45$ is shown as the green (orange) line as an example, and the optimized control $f_{1(2)}$ for $T=46,47,\ldots, 100$ are shown as grey lines, as a function of the scaled time $\tau=t/T$. It can be seen that the controls can be contained within $1\pm 0.3$. The final concurrence and the number of iterations needed to reach the control goal are also given in Fig.~\ref{fig_ctrl} (c), and an example of the entanglement dynamics for $T=45$ is shown in Fig.~\ref{fig_ctrl} (d). The control employed here can drive the final state very close to the maximally entangled target Bell state, with concurrence above $0.9999$, and this strategy may be employed in situations where the entanglement needs to be generated on demand, with a given runtime.

\section{Noises and entanglement generation}\label{sec_opens}

\begin{figure}
    \centering
    \includegraphics[width=.4\textwidth]{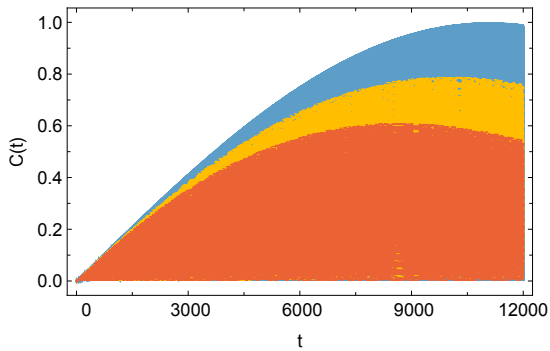}
    \caption{Entanglement dynamics of the magnons as a function of time, the blue lines are under close system dynamics, the yellow dashed lines are obtained from a generic non-Markovian open system treatment, whereas the red dashed lines are obtained from the Markov master equation. It can be seen that the generation entanglement can be quite robust against the noises introduced by the dissipative cavity modes, and the non-Markovian effects may make the entanglement dynamics much less susceptible to dissipation.}\label{fig_openc}
\end{figure}

We have shown how to generate magnon entanglement in various interesting physical settings when the environmental noises can be effectively ignored. In this section, we will study the robustness of generated entanglement when the magnon-cavity systems are under the influence of external noise. Such noise analysis can be realized by using a standard quantum open system approach~\cite{Breuer2002a}. For the physical system under consideration, the remote entanglement generation would require that the two cavities are located in different places, it is of interest to consider the situation where each cavity is coupled to its own environment.

\begin{figure}
    \centering
    \includegraphics[width=.4\textwidth]{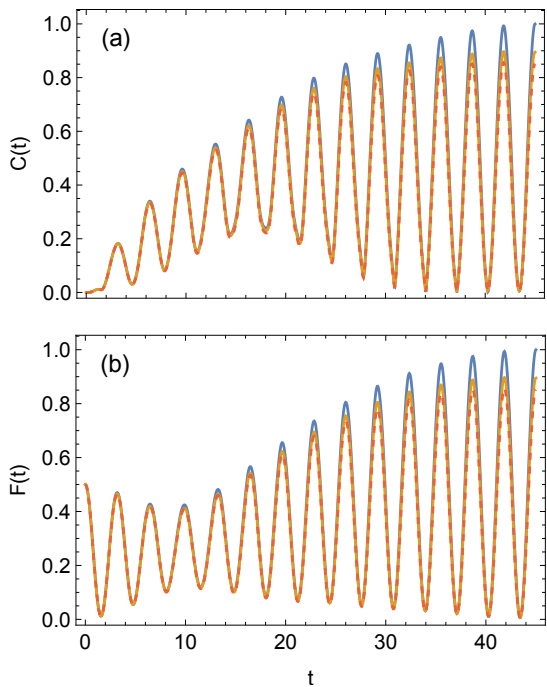}
    \caption{Concurrence (Panel a) and fidelity against the target state (Panel b) of the optimally controlled dynamics under open system for $T=45$, with the same parameters as Fig.~\ref{fig_ctrl}, with system-bath coupling $\lambda=0.1$. The blue line corresponds to the closed system dynamics, the yellow line is obtained under a non-Markovian dynamics while the red dashed line is obtained under the Markov approximation. It can be seen that for the parameters considered here, the control is quite robust against the environment noises, and the concurrence is still able to reach $0.89$ under non-Markovian dynamics while it's slightly lower under Markov approximation at $0.85$.}\label{fig_ctrl_open}
\end{figure}

To be more specific, the noise effect may be described by a bosonic heat bath with Lorentzian spectrum and note that the other types of noises such as classical noises or collective noises can be treated in a similar way \cite{Yu2002c,Yu2003z}. The Hamiltonian $H_{\rm tot}$ for both the system and the heat bath may be written as,
\begin{align}
 H_{\rm tot} &= H_s + H_b + H_{\rm int} \nonumber \\
 &= H_{\rm eff} + \sum_{j=1,2; i} \tilde{\omega}_i B_{j,i} ^\dagger B_{j,i} + \sum_{j=1,2; i} \left(L_j B_{j,i} ^\dagger + h.c.\right),
\end{align}
where $H_s=H_{\rm eff}$ is given in Eq.~\eqref{eq_heff}, $\tilde{\omega}_i$ is the frequency of the $i$-th bath mode, $B_{j,i}(B_{j,i} ^\dagger)$ is the annihilation (creation) operator for the $i$-th mode of bath $j$, and $L_j$ describes the system-bath coupling between cavity $j$ and bath $j$. Here, we consider dissipative microwave and optical cavities,

\begin{align}
 L_j = \lambda_a a_j + \lambda_b b_j
\end{align}

The open system dynamics of the cavity-magnon system coupled to a generic non-Markovian noise is our major concern. The reduced density operator $\rho_s = \mathrm{Tr}_b[\rho_{\rm tot}]$ can be conveniently simulated by using the quantum state diffusion (QSD) equations~\cite{Diosi1998a,Yu1999a,Strunz1999a}. By projecting the bath modes onto the Bargmann coherent states $|z \rangle_j = \exp[z_k B_{j,k}^\dagger]|0 \rangle$, and denoting the wave function of the cavity-magnon system $|\psi_t(z_1^*,z_2^*) \rangle = \langle z_1|\langle z_2|\Psi \rangle$, one can show that the wave function $|\psi_t(z_1^*,z_2^*) \rangle$ satisfies the following stochastic Schr\"{o}diner equation,

\begin{align}
 &\partial_t |\psi_t(z_1^*,z_2^*) \rangle = \nonumber \\
 & \left[-i H_s + L_1 z_1^* + L_2 z_2^* - L_1^\dagger \bar{O}_1 - L_2^\dagger \bar{O}_2 \right]|\psi_t(z_1^*,z_2^*) \rangle,
\end{align}
where $\bar{O}_i = \int ds \alpha_i(t,s) \frac{\delta}{\delta z_{i,s}^*}$ denotes the functional derivative of the state with respect to the noise $z_{i,s}$ and $\alpha_i(t,s)=\mathcal{M}[z_{i,t}z_{i,s}^*]$ is the correlation or memory function of the bath, where $\mathcal{M}[\cdot]$ denotes the ensemble average for the stochastic process. The reduced density matrix is then recovered from the ensemble average $\rho_s = \mathcal{M}[|\psi_t(z_1,z_2)\rangle\langle\psi_t(z_1^*,z_2^*)|]$. One nice feature of the QSD approach is that it is able to encapsulate the environment memory effect into the $\bar{O}$ operator and thus give a time-convolutionless differential equation for the system dynamics. 

The solutions to the open system problems are dependent on the choice of an approximate $\bar{O}$ operator~\cite{Diosi1998a,Yu1999a,Strunz1999a,Jing2010s,Luo2015a}. When the system-bath coupling is not in a strong coupling regime, a leading-order master equation may be derived in the form of~\cite{Jing2015w,Yu2004i}
\begin{align}
 \frac{d}{d t} \rho_s(t)&=-i \left[H_s,\rho_s\right] \nonumber \\
 &+ \sum_{j=1,2} \left[L_j,\rho_t\bar{O_j}^{(0)\dagger}(t)\right]-\left[L_j ^\dagger,\bar{O_j}^{(0)}(t)\rho_t\right], \label{eq_meq}
\end{align}
where the $\bar{O}$ operator is approximated by a noise-free version obtained from
\begin{align}
 \partial_t O_j^{(0)}(t,s)&=\left[-iH_s - \sum_{k=1,2} L_k ^\dagger \bar{O}_k^{(0)}(t),O_j^{(0)}(t,s)\right], \nonumber \\
 \bar{O}_j^{(0)}(t) &= \int_0^t ds \alpha_j(t,s) O_j^{(0)}(t,s).
\end{align}

Here we consider two identical baths with the Ornstein-Uhlenbeck noise~\cite{Diosi1998a,Yu1999a,Strunz1999a} with $\alpha_{1,2}(t,s) = \gamma \exp \left(-\gamma |t-s|\right)/2$, where $1/\gamma$ describes the memory time of the bath. In the limit of $\gamma \rightarrow \infty$, we get the white noise case and the dynamics would be memory-less (Markov). In this case the correlation function $\alpha(t,s)$ becomes a $\delta$-function and the $\bar{O}_j$ operator would be replaced by $L_j/2$. For a finite $\gamma$, we have a non-Markovian bath with memory effects. Taking $\lambda_a=\lambda_b=0.01$, $\gamma = 0.7$, $\omega_m=\omega_b=1$, $\omega_a=1200$, $j_a=90$, $g_m=1.3$ and $g_c=1.5$, we plot the concurrence as a function of time in Fig.~\ref{fig_openc}, where the blue lines are from closed system dynamics for comparison, the concurrence obtained from non-Markovian master equation Eq~\eqref{eq_meq} is plotted in yellow, while the Markovian limit result is plotted in red. It can be seen that for the parameters considered here, the generated entanglement is quite robust against the non-Markovian open system effects, while it drops quite a bit and does not revive in the Markov limit, where the memory effects of the bath are ignored.

It is also of interest to see how environmental noises affect entanglement generation when the optimal control is in action. For the convenience of comparison, we calculate the open system entanglement dynamics for a runtime of $T=45$, shown in Fig.~\ref{fig_ctrl_open} (a), and the fidelity of the evolved state in the open system against the target Bell state is shown in Panel (b) in Fig.~\ref{fig_ctrl_open}. It can be seen that the controlled entanglement scheme remains robust in the presence of the environment noises, where the final concurrence may reach to $0.89$ under non-Markovian noise and $0.85$ under the Markov approximation. Expectedly, when the system-bath couplings become stronger, the generated entanglement is to drop more rapidly. In this case, to maintain robust entanglement one must employ an active control to decouple the system from the detrimental effects~\cite{Jing2015a,Viola1999a,Viola1998s}. 
%We leave the systematic discussions of this issue to a future publication.

\section{Conclusions}\label{sec_conclu}

We considered the remote entanglement generation in a coupled magnon cavity system with an auxiliary frequency conversion system. It is shown that the remote entanglement between the magnon modes can be achieved in several physically interesting settings.

Due to the differences in the microwave and optical frequencies, the wave function components corresponding to the magnon system can be approximated by a superposition of oscillations of slightly different frequencies and thus shows a beat pattern with a periodic envelope. 
As such, the entanglement dynamics also displays a beat pattern. We also show how to use the Krotov optimal approach to generate a desirable entangled state in a prescribed time. For both controlled and uncontrolled cases, we have studied the robustness of the generated entanglement under the influence of an environmental noise modeled by a non-Markovian process in the framework of quantum open systems. We found that the magnon entanglement can be robust against dissipative microwave and optical cavities under the influences of open system effects. Interestingly, we show that the non-Markovian dynamics preserves the entanglement more effectively than the memory-less Markov case. The robust entanglement between remote macroscopic magnon modes is of interest in many quantum information processing tasks such as high precision measurement and detections~\cite{Xia2020s,Khalid2021s,Huang2016t,Deng2021s,Giovannetti2011s,Wang2018f,Pang2014v}.

\begin{acknowledgments}
    This work is supported by the ART020-Quantum Technologies Project. We acknowledge partial support from the U.S. Army under Contact No. W15QKN-18-D-0040.
\end{acknowledgments}
 
% \pagebreak

\onecolumngrid
\appendix

\section{Approximate analytical expressions for the envelopes for the entanglement dynamics}\label{sec_apnx}

In this appendix, we show the approximate analytical expressions for the envelopes for the entanglement dynamics, as displayed in Fig.~\ref{fig_cm_e}. Let
% \begin{widetext}
\begin{align}
\phi_1 &= \cos ^2\left(\frac{g_c^2 j_a t}{\omega_a^2}\right) \sin ^4\left(\frac{g_c^4 j_a t}{2 g_m \omega_a^3}\right) \csc \left(\frac{g_c^2 j_a t \left(g_c^2-2 g_m \omega_a\right)}{2 g_m \omega_a^3}\right) \csc \left(\frac{g_c^2 j_a t \left(g_c^2+2 g_m \omega_a\right)}{2 g_m \omega_a^3}\right) 
\end{align}
% \end{widetext}
defined in the region where
% \begin{widetext}
\begin{align}
 & \cos \left(\frac{g_c^4 j_a t}{g_m \omega_a^3}\right)=\cos \left(\frac{2 g_c^2 j_a t}{\omega_a^2}\right) \text{ or}\nonumber \\
 & \left| \left[\csc \left(\frac{g_c^2 j_a t \left(g_c^2-2 g_m \omega_a\right)}{2 g_m \omega_a^3}\right)-\csc \left(\frac{g_c^2 j_a t \left(g_c^2+2 g_m \omega_a\right)}{2 g_m \omega_a^3}\right)\right] \sin \left(\frac{g_c^2 j_a t}{\omega_a^2}\right)\right| >2
\end{align}
% \end{widetext}
is \emph{not} satisfied. Further, define

\begin{align}
 \phi_2 &= \sin\left(\frac{g_c^2 j_a t}{\omega_a^2}\right) \cos ^2\left(\frac{g_c^4 j_a t}{4 g_m \omega_a^3}\right) \\
 \phi_3 &= \sin\left(\frac{g_c^2 j_a t}{\omega_a^2}\right) \sin ^2\left(\frac{g_c^4 j_a t}{4 g_m \omega_a^3}\right) \\
 \Phi &= \text{max}\left[\cos ^2\left(\frac{g_c^4 j_a t}{4 g_m \omega_a^3}\right),\sin ^2\left(\frac{g_c^4 j_a t}{4 g_m \omega_a^3}\right)\right].
\end{align}

The envelope $ev1$ can then be given by $\sqrt{|\phi_1|}/2$ in the region where $\phi_1$ is defined. The envelope $ev2$ is given by $\text{min}[{|\phi_2|,|\phi_3|}]$ in the region where $\phi_1$ is defined. In the region where $\phi_1$ is not defined, the envelope $ev3$ is given by ${|\phi_2|}$ and the envelope $ev4$ is given by ${|\phi_3|}$. 

\begin{figure}
    \centering
    \includegraphics[width=.45\textwidth]{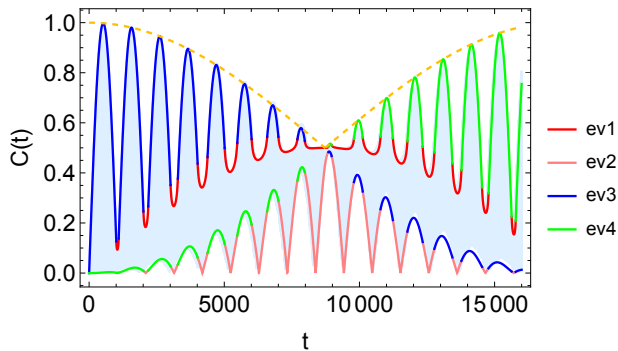}
    \caption{Concurrence between the magnons as a function of time, with the analytical envelopes $ev1 \ldots ev4$. The light-blue lines are the fast-oscillations entanglement, and different colors of the envelope correspond to branches in the envelope. There also exists an overall envelope to the envelopes $ev1 \ldots ev4$, plotted as the orange dashed line.}\label{fig_env4}
\end{figure}

For the parameters and time-range considered in Fig.~\ref{fig_cm_e}, only $ev3$ and $ev4$ are defined. To see the all possible branches of the envelope, we take $\omega_b=\omega_m=1$, $\omega_a=1200$, $g_m=1$, $g_c=12$ and $j_a=30$, and plot the concurrence as a function of time along with the envelopes in Fig.~\ref{fig_env4}. There also exists an overall envelope to the envelopes $ev1 \ldots ev4$, given by $\Phi$, shown as the orange dashed line in Fig.~\ref{fig_env4}.

%%%

\twocolumngrid

%%%%%%%%%%%%%%%%%%%%%%%%%%%%%%%%%%%%%

% \bibliographystyle{prs}
% \bibliography{magn_cc}

%%%%%%%%%%%%%%%%%%%%%%%%%%%%%%%%%%%%%

%%%%%%%%%%%%%%%%%%%%%%%%%%%%%%%%%%%%%

\end{document}